\documentclass[reprint,sort&compress,longbibliography,superscriptaddress,amsmath,aps,showpacs,pra]{revtex4-2}
\usepackage[vietnamese=nohyphenation]{hyphsubst}
\usepackage[vietnamese,british]{babel}
\usepackage{graphicx,epsfig}
\usepackage{float}
\usepackage{hyperref}
\usepackage{breqn}
\usepackage[utf8]{inputenc}
\usepackage{amsmath}
\usepackage{subfigure}
\usepackage[normalem]{ulem}
\usepackage{xcolor}
\usepackage{soul}
\hypersetup{ 
     colorlinks   = true,
     citecolor    = blue,
     urlcolor     = blue
}
\usepackage{physics}
\usepackage{amssymb}
\usepackage{tikz}
\usepackage{feynmf}

\begin{document}

\title{Superconductivity from electronic interactions and spin-orbit enhancement in bilayer and trilayer graphene.}
\author{Alejandro Jimeno-Pozo}
\email{alejandro.jimeno@imdea.org}
\affiliation{$Imdea\ Nanoscience,\ Faraday\ 9,\ 28049\ Madrid,\ Spain$}
\author{H\'ector Sainz-Cruz}
\affiliation{$Imdea\ Nanoscience,\ Faraday\ 9,\ 28049\ Madrid,\ Spain$}
\author{Tommaso Cea}
\affiliation{$Imdea\ Nanoscience,\ Faraday\ 9,\ 28049\ Madrid,\ Spain$}
\affiliation{$Department\ of\ Physical\ and\ Chemical\ Sciences,\ University\ of\ L'Aquila,\ via\ Vetoio,\ Coppito,\ 67100\ L'Aquila,\ Italy$}
\author{Pierre A.\ Pantale\'on}
\affiliation{$Imdea\ Nanoscience,\ Faraday\ 9,\ 28049\ Madrid,\ Spain$}
\author{Francisco Guinea}
\affiliation{$Imdea\ Nanoscience,\ Faraday\ 9,\ 28049\ Madrid,\ Spain$}
\affiliation{$Donostia\ International\  Physics\ Center,\ Paseo\ Manuel\ de\ Lardizabal\ 4,\ 20018\ San\ Sebastian,\ Spain$}
\affiliation{$Ikerbasque\  Foundation, Maria\ de\ Haro\ 3,\ 48013\ Bilbao,\ Spain$}
\date{\today}

\begin{abstract}
We discuss a Kohn-Luttinger-like mechanism for superconductivity in Bernal bilayer graphene and rhombohedral trilayer graphene.\ Working within the continuum model description, we find that the screened long-range Coulomb interaction alone gives rise to superconductivity with critical temperatures that agree with experiments.\ We observe that the order parameter changes sign between valleys, which implies that both materials are valley-singlet, spin-triplet superconductors.\ Adding Ising spin-orbit coupling leads to a significant enhancement in the critical temperature, also in line with experiment, and the superconducting order parameter shows locking between the spin and valley degrees of freedom.
\end{abstract}

\maketitle

{\it Introduction.} Recent experiments report cascades of correlated phases in Bernal bilayer graphene (BBG)~\cite{zhou2022isospin,barrera2022cascade,seiler2022quantum}.\ One of them is spin-polarized superconductivity \cite{zhou2022isospin} with a critical temperature of $T_{c}\approx 26$ mK for fillings near the hole-doped van Hove singularity (vHs) when an out-of-plane electric field together with an in-plane magnetic field are applied on the system.\ Moreover, the authors of Ref.\ \cite{zhang2022spin} have assembled a heterostructure in which a transition metal dichalcogenide (TMD) and BBG enter in synergy, making superconductivity appear over a broader range of electron filling and magnetic field, even without the latter, and enabling a striking increment in critical temperature to $T_{c}\approx$ 260 mK, an effect that is attributed to Ising spin-orbit coupling (SOC).\ The stability and structural simplicity of BBG are clear advantages for experimental reproducibility, a major obstacle in Moiré materials \cite{lau2022reproducibility}, in which all samples differ due to angle disorder \cite{uri2020, schapers2022raman} and strains \cite{kazmierczak2021strain}.\ A previous experiment showed that rhombohedral trilayer graphene (RTG) is a superconductor as well \cite{zhou2021superconductivity} and several theories about it have been proposed \cite{chou2021acousticRTG,dai2021mott,dong2021superconductivity,ghazaryan2021unconventional,chatterjee2021inter,cea2022superconductivity,you2022kohn,szabo2022metals,qin2022functional,dai2022quantum,lu2022correlated}.\ A key piece in the puzzle of graphene superconductors is that they all display cascades of flavour-symmetry-breaking phase transitions \cite{zhou2022isospin,zondiner2020cascade,wong2020cascade,Zhou2021HalfQuarterMetals,barrera2022cascade,seiler2022quantum}, perhaps hinting at a common origin of superconductivity.

In these materials, and in BBG in particular, the proximity of superconductivity to flavour-polarized metallic phases and its Pauli limit violation point to an unconventional spin-triplet pairing mediated by electrons.\ In Ref.\ \cite{Szabo2022Bilayer}, a model is discussed in which short-range momentum-independent interactions and proximity to symmetry broken phases induce spin-triplet $f$-wave \mbox{pairing}.\ The authors of Ref.\ \cite{dong2022spin} propose a superconductivity from repulsion mechanism, in which the Coulomb interaction dressed by soft quantum-critical modes drives pairing.\ In their model, the magnetic field induces spin imbalance, which makes the interaction acquire a dependence on frequency or on soft-mode momenta, leading to valley-singlet, spin-triplet \textit{s}-wave or valley-triplet spin-triplet \textit{p}-wave pairings, respectively.\ In Refs.\ \cite{chou2022acoustic,chou2022enhanced}, a pairing mechanism mediated by acoustic phonons is investigated, which is compatible with spin-singlet \textit{s}-wave and spin-triplet \textit{f}-wave pairings.\ However, the authors assume that the Coulomb interaction is detrimental for superconductivity.\ In contrast, we find that the Coulomb interaction alone enables superconductivity, see also \cite{dong2022spin}.

Here, we present a framework in which superconductivity in BBG emerges only from the long-range Coulomb interaction, arguably the simplest explanation.\ Due to screening by particle-hole pairs, this repulsive interaction between electrons becomes attractive and leads to pairing.\ We use a diagrammatic technique similar to the Kohn-Luttinger approach \cite{Kohn1965}, which we have already applied to twisted bilayer and trilayer graphene \cite{cea21Coulomb,phong2021band}, as well as to RTG \cite{cea2022superconductivity}, thus allowing for a direct comparison of superconductivity in these systems and in BBG.\ We find superconductivity with critical temperature comparable to the experimental one \cite{zhou2022isospin}, near the vHs.\ Including Ising SOC leads to significant increments in the critical temperature, as seen experimentally \cite{zhang2022spin}.\ We observe that the order parameter (OP) changes sign within each valley.\ Adding a short-range Hubbard repulsion, we conclude that the OP also changes sign between valleys, showing that BBG is a spin-triplet superconductor.\ The rest of the paper is organized as follows: first we describe the continuum model and the Kohn-Luttinger-like framework for superconductivity.\ Then we present and discuss the results of the critical temperature, the superconducting OP and the positive effect of adding SOC.
\begin{figure}[h]
    \includegraphics[width=8.5cm]{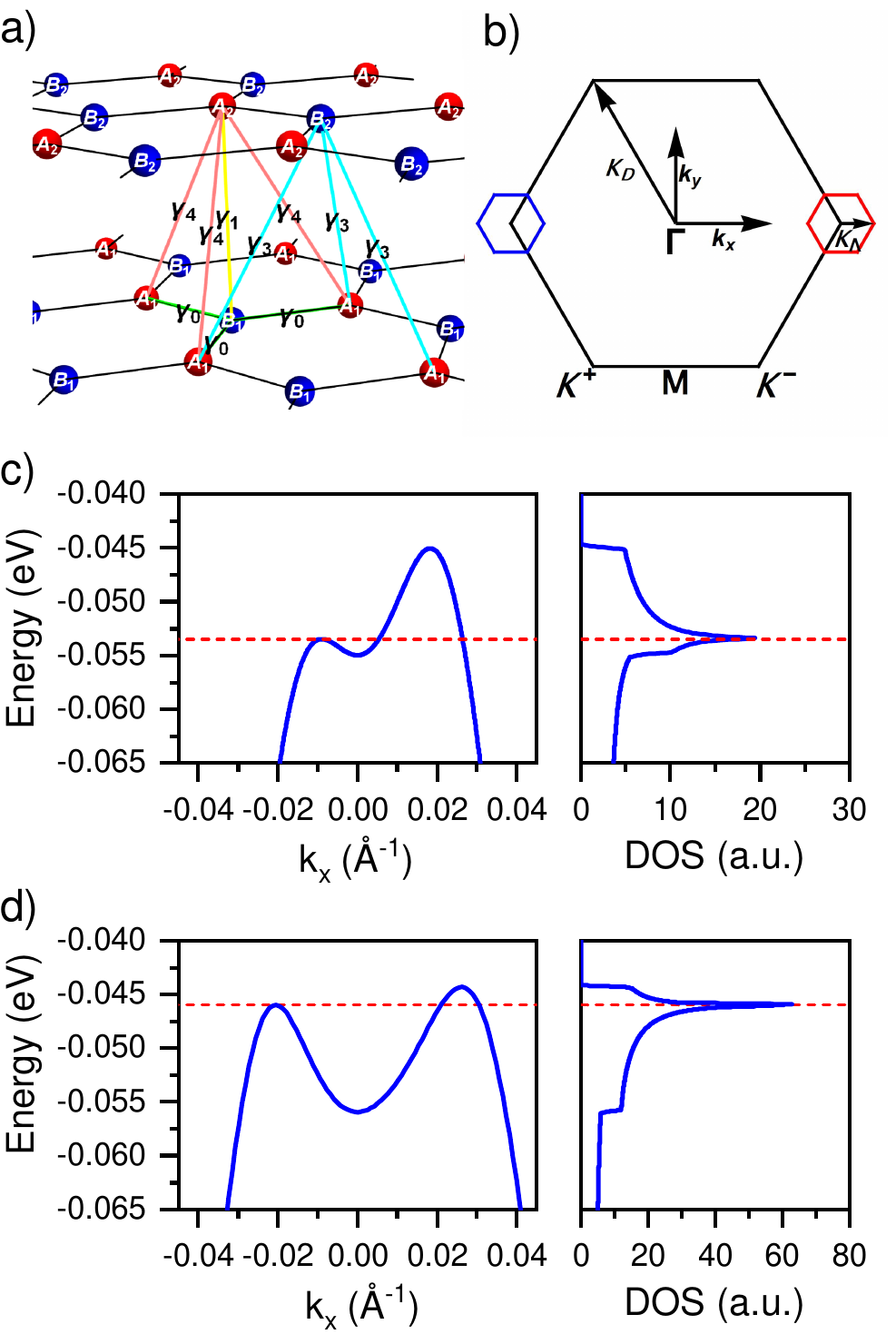}
    \caption{(a) Crystal lattice and hoppings of BBG.\ (b) Brillouin zone of BBG and RTG.\ We perform the calculation of superconductivity considering electron states in hexagonal grids around the Dirac points, with ultraviolet momentum cutoff $k_\Lambda = 0.025 K_{D}$ for BBG (and $0.035 K_{D}$ for RTG), with $K_D=4\pi/3a$ the Dirac point modulus, which contain the Fermi surface and where the continuum model is a good approximation.\ Red and blue hexagons are plotted with $k_\Lambda = 0.2 K_{D}$ for clarity.\ (c, d) Continuum model valence band and DOS of BBG and RTG with electric-field-induced gaps of $98$ and $74$ meV, respectively \cite{fields}.\ Dashed red lines mark the energies of the vHs.}
    \label{fig:1}
\end{figure}

{\it The continuum model.} Bernal bilayer graphene (BBG) refers to two stacked graphene layers so that atoms belonging to sublattice A of layer 1 lie over the atoms in sublattice B of layer 2.\ Similarly to monolayer graphene, BBG is a semi-metal in which the low energy bands touch at the Dirac points, but with parabolic instead of linear dispersion \cite{mccann2006landau,novoselov2006unconventional}.\ This band touching makes the charge susceptibility and other susceptibilities diverge \cite{nilsson2006electron}, which leads to symmetry-broken phases \cite{Zhang2011,min2008pseudospin,vafek2010interacting,vafek2010many,nandkishore2010dynamical,lemonik2012competing,zhang2012distinguishing,cvetkovic2012electronic, martin2010local}, as seen in experiments \cite{weitz2010broken, mayorov2011interaction,velasco2012transport,freitag2012spontaneously,ju2015topological,li2017even, zibrov2017tunable}.\ A perpendicular electric field opens a gap and the bands acquire a `Mexican hat' profile \cite{zhang2009direct,mccann2013electronic}, as shown in Fig.\ \ref{fig:1}(c,d), making the density of states diverge logarithmically near the band edges, causing vHs and setting the stage for new correlated phenomena \cite{maher2014tunable,lee2014chemical,kou2014electron,geisenhof2021quantum,zhou2022isospin,barrera2022cascade,seiler2022quantum,zhang2022spin}.

In the continuum approximation the hamiltonian of BBG can be expressed in the basis $\{ \Psi_{A_{1}}, \Psi_{B_{1}},  \Psi_{A_{2}}, \Psi_{B_{2}} \}$ as \cite{mccann2013electronic},
\begin{equation}
    \mathcal{H}_{\xi} = \mqty( V/2 & v_{0}\pi^{\dagger} & -v_{4}\pi^{\dagger} & v_{3}\pi \\ v_{0}\pi & V/2 + \Delta^{\prime} & \gamma_{1} &  -v_{4}\pi^{\dagger} \\ -v_{4}\pi & \gamma_{1} & -V/2 + \Delta^{\prime} & v_{0}\pi^{\dagger} \\ v_{3}\pi^{\dagger} & -v_{4} \pi & v_{0} \pi &  -V/2 ),
    \label{eq:1}
\end{equation}
where $\pi = \xi p_{x} + \mathrm{i} p_{y}$, $\xi=\pm1$ is the valley index and $\vb{p} = \vb{k} - \vb{K}^{\xi}$ is the momentum with respect to the $\vb{K}^{\xi}$ Dirac point, $v_{i} = \frac{\sqrt{3} a \gamma_{i}}{2\hbar}$ are effective velocities, $a$ is the lattice constant and $\gamma_{i}$ the different hoppings in the system, shown in Fig.\ \ref{fig:1}(a).\ $V$ corresponds to an applied electric field and $\Delta^{\prime}$ is the energy difference between dimer ($A_1$, $B_2$) and non-dimer ($A_2$, $B_1$) sites.\ In Ref.\ \cite{SM} we list the values \cite{mccann2013electronic, miscBBG}, and describe the hamiltonian of RTG.

{\it Kohn-Luttinger-like superconductivity.} We propose that electronic interactions alone are enough to induce superconductivity in BBG and RTG and that the pairing glue is the screened long-range Coulomb potential $V_{scr}$.\ We employ a similar approach to the Kohn-Luttinger theory \cite{Kohn1965}, following an analogous method as in \cite{cea2022superconductivity}.\ We use the Random Phase Approximation to take into account the screened direct interaction to infinite order, while neglecting the contribution from the exchange interaction.\ The multiplicity of the direct diagrams equals the number of flavours, $N_f=4$, so the approximation can be considered an expansion in powers of $1/N_f$.\ Under these assumptions the screened interaction is given by, 
\begin{equation}
    V_{scr}(\vb{q}) = \frac{V_{C}(\vb{q})}{1 - \Pi(\vb{q})V_{C}(\vb{q})},
    \label{eq:2}
\end{equation}
where $V_{C}(\vb{q}) = \frac{2\pi e^2}{\epsilon |\vb{q}| }\tanh{\left(d |\vb{q}|\right)}$ is the bare Coulomb potential, with $e$ the electron charge, $\epsilon=4$ the dielectric constant associated hBN encapsulation and $d=40$ nm the distance to the metallic gate.\ $\Pi(\vb{q})$ corresponds to the zero-frequency limit of the charge susceptibility,
\begin{equation}
    \Pi(\vb{q}) = \frac{N_f}{\Omega}\sum_{\vb{k},m,n} \frac{f(\epsilon_{n,\vb{k}}) - f(\epsilon_{m,\vb{k}+\vb{q}})}{\epsilon_{n,\vb{k}} - \epsilon_{m,\vb{k}+\vb{q}}} \abs{\braket{\Psi_{m,\vb{k}+\vb{q}}}{\Psi_{n, \vb{k}}}}^{2},
    \label{eq:3}
\end{equation}
where $m,n$ are band indexes, $f$ is the Fermi-Dirac distribution and $\epsilon=E-\mu$ with $\mu$ being the Fermi energy, E are the eigenvalues, $\Psi$ the eigenvectors of (\ref{eq:1}) and $\Omega$ is the area of the system.\ 
\begin{figure*}[t!]
    \centering
    \includegraphics[width=\textwidth]{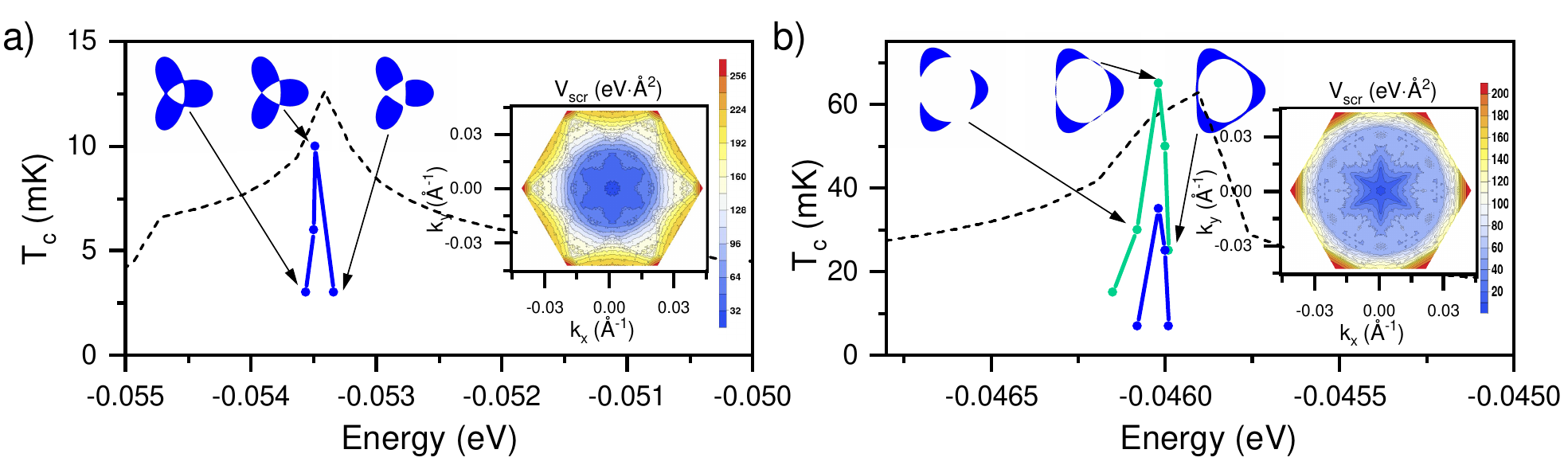}
    \caption{Superconducting critical temperature due to only the screened long-range Coulomb interaction (blue), and including also the short-range interaction (green), and DOS (a.u., dashed black) versus Fermi energy in (a) BBG with hole doping, near the vHs ($n_e\approx-0.87\cdot 10^{12}$ cm$^{-2}$), we observe a $T_{c}\approx10$ mK, and (b) RTG with hole doping, near the vHs ($n_e\approx-0.6\cdot 10^{12}$ cm$^{-2}$), $T_{c}\approx33$ mK and it raises to $T_{c}\approx65$ mK after including the short-range interaction.\ The top insets show the shape of the Fermi surfaces, which experience a change in topology at the vHs due to Lifshitz transitions.\ Right insets: screened Coulomb potential to $k_{\Lambda}$ at the energy for which $T_{c}$ is maximum.}
    \label{fig:2}
\end{figure*}
Once we perform this calculation within each valley, we  include a short-range, repulsive Hubbard $U=3$ eV \cite{Wehling2011,Wehling2013}, which allows electrons to exchange valley and fixes the sign of the order parameter in each valley.\ We define the dimensionless anomalous expectation values in both valleys $\tilde{\Delta}^{+, i, j}(\vb{k}) = \expval{c_{\vb{k}, i, \vb{K}^{+}, \uparrow}^{\dagger} c_{-\vb{k}, j, \vb{K}^{-}, \downarrow}^{\dagger}}$ and $\tilde{\Delta}^{-, i, j}(\vb{k}) = \expval{c_{\vb{k}, i, \vb{K}^{-}, \uparrow}^{\dagger} c_{-\vb{k}, j, \vb{K}^{+}, \downarrow}^{\dagger}}$.\ Therefore the gap equation is given by, 
\begin{equation}
    \begin{split}
    \tilde{\Delta}^{+, i,j}\left(\vb{k}\right) &= -\frac{K_{B}T}{\Omega}\sum_{\vb{q}, \omega}\sum_{i^{\prime},j^{\prime}} \\& V_{scr}\left(\vb{k} - \vb{q}\right) G_{\vb{K}^{+}}^{i,i^{\prime}}\left(\vb{q},\mathrm{i}\omega\right)G_{\vb{K}^{-}}^{j,j^{\prime}}\left(-\vb{q},-\mathrm{i}\omega\right)\tilde{\Delta}^{+,i^{\prime},j^{\prime}}\left(\vb{q}\right) \\&+ U G_{\vb{K}^{-}}^{i,i^{\prime}}\left(\vb{q},\mathrm{i}\omega\right)G_{\vb{K}^{+}}^{j,j^{\prime}}\left(-\vb{q},-\mathrm{i}\omega\right)\tilde{\Delta}^{-, j^{\prime},i^{\prime}}\left(-\vb{q}\right),
    \end{split}
    \label{eq:4}
\end{equation}
where indexes $i,j$ denote layer and sublattice degrees of freedom, $\omega$ is the Matsubara frequency and $G_{K^{\xi}}^{i,j}\left(\vb{q},\mathrm{i}\omega\right)$ are electronic Green's functions.\ Eq.\ (\ref{eq:4}) can be understood in terms of the convolution of the gap with a kernel,
\begin{equation}
    \tilde{\Delta}(\vb{k}) = \mqty(\tilde{\Delta}^{+}(\vb{k}) \\ \tilde{\Delta}^{-}(\vb{k})) = \sum_{\vb{k}^{\prime}}\mqty(\Gamma^{+}(\vb{k},\vb{k}^{\prime}) & \tilde{U}(\vb{k},\vb{k}^{\prime})\\ \tilde{U}^{\dagger}(\vb{k},\vb{k}^{\prime}) & \Gamma^{-}(\vb{k},\vb{k}^{\prime}))\tilde{\Delta}(\vb{k}^{\prime}), 
    \label{eq:5}
\end{equation}
so that the solution of this gap equation is achieved when the maximum eigenvalue of the hermitian kernel reaches a value of 1, indicating the $T_{c}$ at which the scattering amplitude for pairs of carriers of opposite momenta and energy within the Fermi surface has a pole.\ The corresponding eigenvector is the superconducting order parameter.\ The long-range Coulomb kernel is,
\begin{equation}
\hspace*{-0.5cm} 
    \Gamma_{m,n}^{\xi}(\vb{k},\vb{k^{\prime}})= -\frac{V_{scr}(\vb{k}-\vb{k^{\prime}})}{\Omega} S^{\xi,\xi}_{m,n}(\vb{k},\vb{k^{\prime}}) R^{\xi,\xi}_{m,m}(\vb{k},\vb{k}) R^{\xi,\xi}_{n,n}(\vb{k}^{\prime},\vb{k}^{\prime}),
    \label{eq:6}
\end{equation}
while the short-range kernel is,
\begin{equation}
    \tilde{U}_{m,n}(\vb{k},\vb{k^{\prime}})= -\frac{U}{\Omega} S^{+,-}_{m,n}(\vb{k},\vb{k^{\prime}}) R^{+,-}_{m,m}(\vb{k},\vb{k}) R^{-,+}_{n,n}(\vb{k}^{\prime},\vb{k}^{\prime}),
    \label{eq:7}
\end{equation}
with\\
$R^{\xi,\xi^{\prime}}_{m,n}(\vb{k},\vb{k^{\prime}}) = \sqrt{\left(f(-\epsilon^{\xi}_{m,\vb{k}}) - f(\epsilon^{\xi^{\prime}}_{n,\vb{k}^{\prime}})\right) / \left(\epsilon^{\xi}_{m,\vb{k}} + \epsilon^{\xi^{\prime}}_{n,\vb{k}^{\prime}}\right)}$ and $S^{\xi,\xi^{\prime}}_{m,n}(\vb{k},\vb{k^{\prime}}) = \abs{\braket{\Psi^{\xi}_{m,\vb{k}}}{\Psi^{\xi^{\prime}}_{n,\vb{k^{\prime}}}}}^{2}$.\ Once Eq.\ \eqref{eq:5} is solved, gaps with dimensions of energy can be obtained as, $\Delta^{+}(\vb{k}) = V_{scr}(\vb{k})\tilde{\Delta}^{+}(\vb{k}) + U \tilde{\Delta}^{-}(\vb{k})$ and $\Delta^{-}(\vb{k}) = V_{scr}(\vb{k})\tilde{\Delta}^{-}(\vb{k}) + U\tilde{\Delta}^{+}(\vb{k})$. 

{\it Results.} In Figure \ref{fig:2} we present the superconducting critical temperature of hole-doped BBG and RTG as a function of Fermi energy.\ We obtain a $T_c$ of 10 mK for BBG and 65 mK for RTG, which match well with those observed experimentally \cite{zhou2022isospin, zhou2021superconductivity}.\ In both materials superconductivity survives only in narrow energy intervals around the vHs.\ It is worth noting that in RTG, including the short-range interaction doubles $T_c$, but it does not increase $T_c$ in BBG.\ To obtain $T_c$, we use a $\vb{k}$-point density of up to $\mathcal{O}(10^7)\r{A}^2$ for BBG and $\mathcal{O}(10^6)\r{A}^2$ for RTG.\ Since states close to the Fermi surface give the principal contribution to Eq.\ (\ref{eq:5}), we cut off phase space by considering only states with $\abs{\epsilon_{n,\vb{k}}}\leqslant30$ meV.\ These continuum model results are in good agreement with our previous tight-binding calculation for RTG \cite{cea2022superconductivity}.\ The insets in Fig.\ \ref{fig:2} show the screened potential in momentum space, which is always repulsive, i.e.\ positive, and its value increases with increasing momentum.\ This dependence favors non \textit{s}-wave superconductivity \cite{MC13}.

Moreover, inspired by the authors of Ref.\ \cite{zhang2022spin}, we study the effect of stacking a TMD on top of BBG.\ We assume that the main effect is an induced Ising spin-orbit coupling (SOC) in BBG, which breaks the equivalence between Cooper pairs $\ket{\vb{K}^{+},\uparrow; \vb{K}^{-},\downarrow}$ and $\ket{\vb{K}^{+},\downarrow; \vb{K}^{-},\uparrow}$, so the order parameter cannot be defined as a spin singlet or a spin triplet.\ The Ising hamiltonian which we add to (\ref{eq:1}) is just $\mathcal{H}_{I,\xi,s} = s\xi\lambda_{I}\mathbb{I}$, where s is the spin index.\ This interaction does not reshape the band structure, it just promotes one type of Cooper pair, by lowering the energy of two spin-valley flavours and raising that of the other two, splitting the original vHs into two.\ 
\begin{figure}[t!]
    \centering
    \includegraphics[width=8cm]{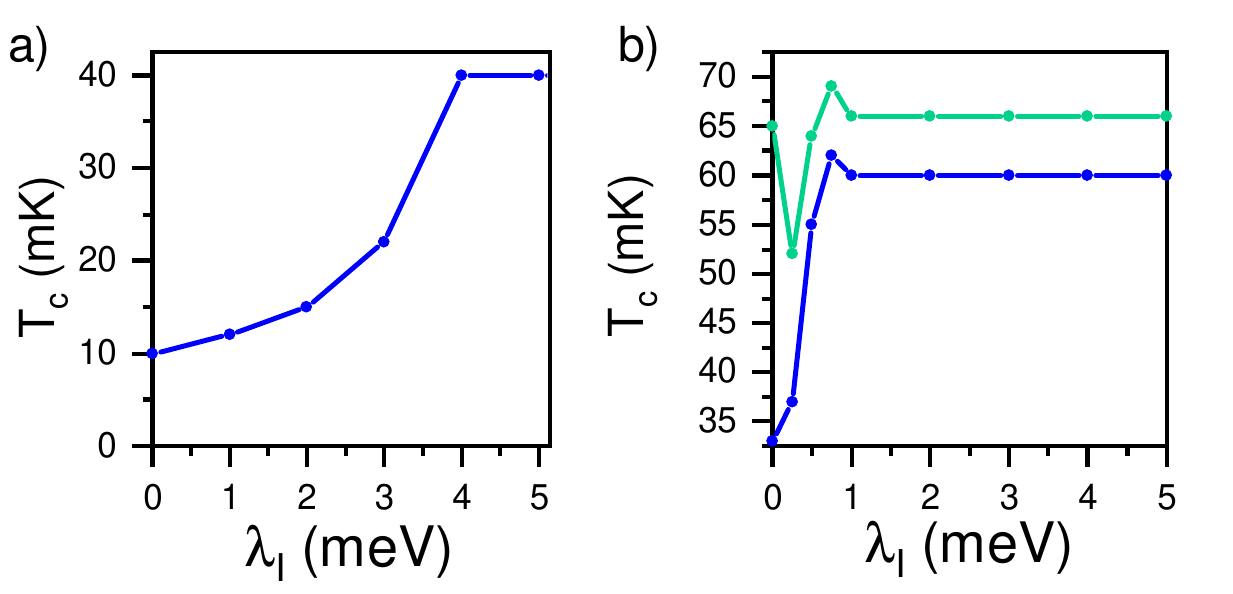}
    \caption{Ising spin-orbit coupling enhancement of critical temperature in hole-doped (a) BBG, for which $T_c$ augments by a factor of 4 and saturates at $\lambda_I=4$ meV. (b) RTG, for which $T_c$ due to long-range interactions (blue) augments by a factor of 2 and saturates at $\lambda_I=1$ meV.\ In contrast, Ising SOC is detrimental to short-range interactions, so $T_c$ does not increase much when they are included (green).}
    \label{fig:3}
\end{figure}
As shown in Figure \ref{fig:3}, Ising SOC has a very positive effect on superconductivity in both materials: it increases $T_c$ by a factor of 4 in BBG, to 40 mK, congruent with the experiment, in which a factor of 10 increment was seen \cite{zhang2022spin} and in RTG it brings forth a comparable increment of $T_c$ due to long-range interactions only.\ The two vHs lead to two superconducting domes as a function of Fermi energy, as observed in Ref.\ \cite{zhang2022spin}, see \cite{SM}.\ It is worth noting that the spin-orbit enhancements of $T_{c}$ saturate once the SOC fully polarizes the non-superconducting state into a half-metal \cite{SM}.

\begin{figure}[t!]
    \centering
    \includegraphics[width=9cm]{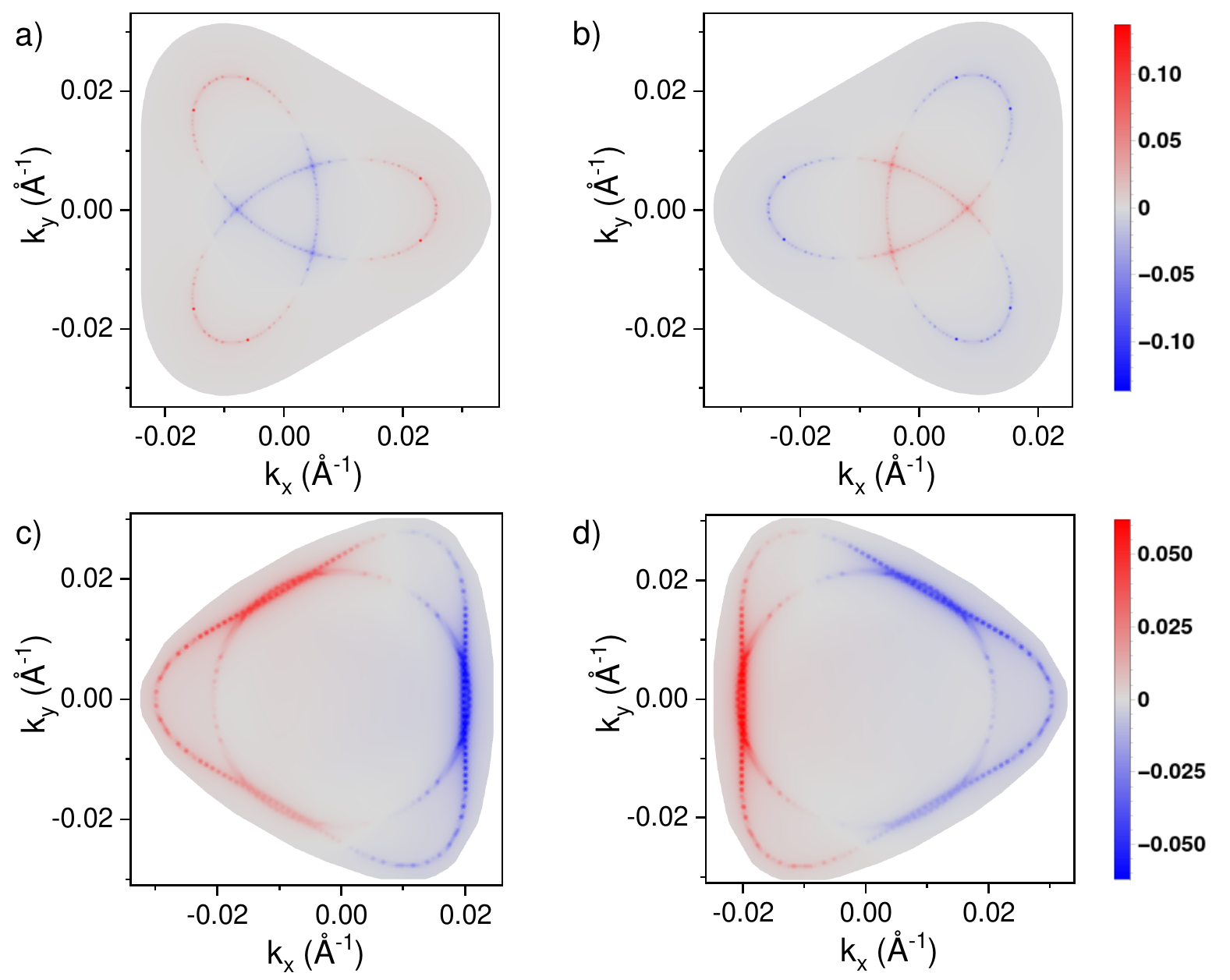}
    \caption{(a) Superconducting order parameter (OP) of BBG in valley $K^{+}$, (b) in valley $K^{-}$, with hole doping and $T_{c}\approx10$ mK.\ Within a single valley, the OP has hotspots along the edges of the Fermi surface and changes sign between the inner and outer edges.\ (c) OP of RTG in valley $K^{+}$, (d) in valley $K^{-}$, with hole doping and $T_{c}\approx65$ mK, showing intensity stripes along the edges of the Fermi surface.\ In both materials, the OP changes sign between valleys, which means that BBG and RTG are valley-singlet, spin-triplet superconductors.}
    \label{fig:4}
\end{figure}

Our model also yields the superconducting order parameters (OPs), shown in Figure \ref{fig:4}.\ The OPs have structure within each valley: in BBG they display hotspots and sign changes along the edges of the Fermi surface.\ The sign is opposite for the inner and outer Fermi edges.\ In RTG, the maximum eigenvalue of the kernel is degenerate, leading to $C_3$-symmetry breaking.\ Such degeneracy is not typical of conventional superconductors and hints at an exotic type of superconductivity.\ We observe that the OPs in both materials change sign between valleys, forming valley-singlets.\ Since the overall electron wavefunction must be antisymmetric, this implies that the pairs are spin-triplets.\ The order parameter that we find in the presence of spin-orbit coupling is similar to the spin-valley locking model in \cite{LPS22}.\ Our results are consistent with the phenomenological model discussed in \cite{curtis2022stabilizing}.

{\it Discussion.} BBG and RTG are 2D superconductors with great structural stability and low disorder.\ These are decisive advantages to understand the physics at play, as they enable very reproducible experiments.\ There is also compelling evidence that both support spin-triplet superconducting phases and hence the pairing is most likely unconventional and mediated by electrons.\ Here, we show that a Kohn-Luttinger-like mechanism for superconductivity suffices to give rise to superconductivity in BBG and RTG, with critical temperatures in good agreement with experiments on both materials \cite{zhou2022isospin, zhou2021superconductivity,zhang2022spin}.\ The screened long-range Coulomb interaction is the sole responsible for pairing.\ At fillings near the vHs, electron interactions strongly screen the Coulomb potential, which glues pairs of carriers, one in each valley, and causes superconductivity.\ Of course, other excitations may dress the Coulomb interaction and contribute to superconductivity in BBG \cite{Szabo2022Bilayer,dong2022spin, chou2022acoustic} and RTG \cite{chou2021acousticRTG,dai2021mott,dong2021superconductivity,ghazaryan2021unconventional,chatterjee2021inter,you2022kohn,szabo2022metals,qin2022functional,dai2022quantum,lu2022correlated}. 

Furthermore, our results support the proposal in Ref.\ \cite{zhang2022spin} that Ising SOC is what causes the enhancement of $T_c$ in BBG when the TMD WSe$_2$ is stacked on top of it.\ Based on the results presented here, we predict that the same idea will boost superconductivity in RTG by a comparable amount as in BBG.\ Since the four times increment we obtain in BBG underestimates the ten times increment observed in the experiment, it is likely that our result for RTG similarly underestimates the SOC enhancement that superconductivity will experience in this material.\ Hence, we conjecture that RTG with WSe$_2$ stacked on top 
will reach $T_c\sim1$ K, putting it on par with magic-angle twisted bilayer graphene and the other twisted stacks \cite{cao2018unconventional, yankowitz2019, lu2019superconductors,park2021tunable, hao2021electric,zhang2021ascendance,park2022robust}.

The OPs reveal two key aspects of superconductivity.\ First, the OPs show sign changes within each valley, which are characteristic of weak coupling superconductivity, in which electrons interact and form pairs with high angular momentum, such as those with \textit{p}-wave or \textit{f}-wave symmetry.\ Another consequence of this intravalley structure is that long-range disorder will suppress superconductivity in these materials, unlike in twisted stacks.\ Second, the sign of the OPs changes between valleys, hence the pairs are spin-triplets.

In BBG and RTG, the screened long-range Coulomb interaction suffices to reach critical temperatures that agree with experiments, while in twisted stacks it is necessary to include Umklapp processes involving large momentum transfers and dressing by phonons, pointing to a rich interplay between interactions and wavefunction complexity \cite{cea21Coulomb, phong2021band}.\ Also, the rigidity of the bands of BBG and RTG leads to narrow superconducing sleeves, because the phase vanishes when the Fermi energy passes the vHs.\ In contrast, in twisted stacks the bands reshape with filling and there is a pinning of the Fermi energy to the vHs \cite{cea2019electronic}, allowing for wide superconducting domes.

In conclusion, we find that superconductivity emerges in BBG and RTG from the screened long-range Coulomb interaction alone.\ This proposal, based on Kohn-Luttinger theory, yields results consistent with three major experimental advances \cite{zhou2022isospin, zhou2021superconductivity,zhang2022spin}.\ We observe critical temperatures that agree with experiments.\ The inclusion of Ising SOC results in prominent increments in critical temperature for both materials, in harmony with the experimental observation in BBG.\ In the absence of Ising SOC the OPs show that both materials are valley-singlet, spin-triplet superconductors, while this classification cannot be applied when SOC is present since it locks the spin and valley degrees of freedom. 

{\it Acknowledgments.} We are thankful to A.\ V.\ Chubukov for illuminating discussions.\ We acknowledge support from the Severo Ochoa programme for centres of excellence in R\&D (Grant No.\ SEV-2016-0686, Ministerio de Ciencia e Innovaci\'on, Spain); from the European Commission, within the Graphene Flagship, Core 3, grant number 881603 and from grants NMAT2D (Comunidad de Madrid, Spain) and SprQuMat (Ministerio de Ciencia e Innovaci\'on, Spain).


%

\newpage

\setcounter{equation}{0}
\setcounter{figure}{0}
\setcounter{table}{0}
\makeatletter
\renewcommand{\theequation}{S\arabic{equation}}
\renewcommand{\thefigure}{S\arabic{figure}}

\onecolumngrid

\textbf{\large Supplementary information for `Superconductivity from electronic interactions and spin-orbit enhancement in bilayer and trilayer graphene'}


\section{The continuum model of BBG and RTG}
Bernal bilayer graphene (BBG) refers to two vertically stacked graphene layers so that atoms belonging to the sublattice A of layer 1 lie over the atoms of the sublattice B of layer 2, while rhombohedral trilayer graphene (RTG) alludes to a stack of three layers in which the atoms of sublattice A of layer 3 are located over the atoms of sublattice B of layer 2 while the atoms of sublattice A of this layer lie over the atoms of sublattice B of layer 1. 

Similar to monolayer graphene, BBG and RTG are semi-metals with band touching at the Dirac points in the low energy region, but with parabolic instead of linear dispersion \cite{mccann2006landau,novoselov2006unconventional}.\ An external perpendicular electric field opens a gap in both systems and the low energy bands nearly flatten around the Dirac points while acquiring `Mexican hat' profiles \cite{mccann2013electronic}, as can be seen in Fig.\ \textcolor{red}{1}(c,d) in the main text.\ The flattening of the low energy bands leads to logarithmic divergences in the densities of states which correspond to van Hove singularities (vHs).

In real space the BBG and RTG unit cell is defined by vectors $\vb{a}_{1} = \frac{a}{2}\mqty(1, \sqrt{3})$ and $\vb{a}_{2} = \frac{a}{2}\mqty(1, -\sqrt{3})$, where $a=2.46\r{A}$ is the graphene lattice constant.\ The Brillouin zone (BZ) of both systems is defined by the reciprocal lattice vectors $\vb{b}_{1} = \frac{2\pi}{a}\mqty(1, \frac{1}{\sqrt{3}})$ and $\vb{b}_{2} = \frac{2\pi}{a}\mqty(1, -\frac{1}{\sqrt{3}})$.\ In the continuum approximation the hamiltonian of BBG can be expressed in the basis $\{ \Psi_{A_{1}}, \Psi_{B_{1}},  \Psi_{A_{2}}, \Psi_{B_{2}} \}$ as follows from  \cite{mccann2013electronic},
\begin{equation}
    \mathcal{H}_{BBG} = \mqty( V/2 & v_{0}\pi^{\dagger} & -v_{4}\pi^{\dagger} & v_{3}\pi \\ v_{0}\pi & V/2 + \Delta^{\prime} & \gamma_{1} &  -v_{4}\pi^{\dagger} \\ -v_{4}\pi & \gamma_{1} & -V/2 + \Delta^{\prime} & v_{0}\pi^{\dagger} \\ v_{3}\pi^{\dagger} & -v_{4} \pi & v_{0} \pi & -V/2 ),
    \label{eq:S1}
\end{equation}
In the case of RTG, the hamiltonian in the basis $\{\Psi_{A_{1}}, \Psi_{B_{1}}, \Psi_{A_{2}}, \Psi_{B_{2}}, \Psi_{A_{3}}, \Psi_{B_{3}} \}$ takes the following form, see \cite{Zhang2010},
\begin{equation}
    \mathcal{H}_{RTG} = \mqty( V + \Delta_{2} + \delta & v_{0}\pi^{\dagger} & v_{4}\pi^{\dagger} & v_{3}\pi & 0 & \gamma_{2}/2 \\ v_{0}\pi & V + \Delta_{2} & \gamma_{1} &  v_{4}\pi^{\dagger} & 0 & 0\\ v_{4}\pi & \gamma_{1} & -2\Delta_{2} & v_{0}\pi^{\dagger} & v_{4}\pi^{\dagger} & v_{3}\pi \\ v_{3}\pi^{\dagger} & v_{4}\pi & v_{0}\pi & -2\Delta_{2} & \gamma_{1} & v_{4}\pi^{\dagger} \\ 0 & 0 & v_{4}\pi & \gamma_{1} & -V + \Delta_{2} & v_{0}\pi^{\dagger} \\ \gamma_{2}/2 & 0 & v_{3}\pi^{\dagger} & v_{4}\pi & v_{0}\pi & -V + \Delta_{2} + \delta),
    \label{eq:S2}    
\end{equation}
with $\gamma_{0}=3.1$, eV, $\gamma_{1}=0.38$ eV, $\gamma_{2}=-0.015$ eV, $\gamma_{3}=-0.29$ eV, $\gamma_{4}=-0.141$ eV, $\delta = -0.015$ eV and $\Delta_{2}=-0.0023$ eV, extracted from \cite{Zibrov2018,Zhou2021HalfQuarterMetals}.
\begin{figure}[h]
    \centering
    \includegraphics[width=\textwidth]{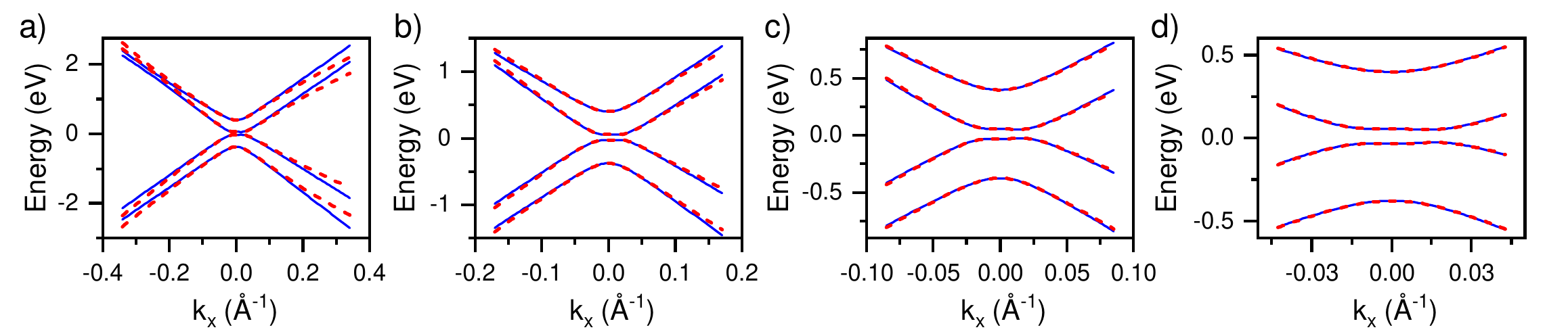}
    \caption{Comparison between tight-binding (dotted red) and continuum model (solid blue) low energy band structure of BBG near the Dirac point for (a) $k_\Lambda=0.2$, (b) $k_\Lambda=0.1$, (c) $k_\Lambda=0.05$ and (d) $k_\Lambda=0.025$, in units of $K_D$.}
    \label{fig:comparison_TB_cont}
\end{figure}
In order to set an ultraviolet momentum cutoff $k_\Lambda$, we compare the low energy band structure of BBG within the tight-binding and continuum models, which can be seen in Fig.\ \ref{fig:comparison_TB_cont}, Ref.\ \cite{cea2022superconductivity} shows a similar comparison for RTG.\ We observe that for $k_\Lambda \le 0.05 K_D$, where $K_D=4\pi/3a$ is the modulus of the Dirac point momentum, both models are in good agreement.\ The idea is to choose the lowest possible $k_\Lambda$, such that there is a high density of $\vb{k}$-points near the Fermi surface, and the energy resolution of the calculation allows for the determination of the critical temperature even if it is low. To find a convergent $T_c$, an energy resolution of $K_BT_c$ is required, where $K_B$ is the Boltzmann constant.\
However, there is a lower bound on $k_\Lambda$.\ This is because superconductivity comes from kernel matrix elements that involve scatterings with momentum exchange $\vb{q}=\vb{k-k'}$, in which $\vb{k}$ and $\vb{k'}$ are within the Fermi surface.\ Therefore, it is necessary to set  $k_\Lambda>max{\mid \vb{q} \mid}$, in order to calculate the screened potential for all relevant momentum values.\ For the displacement fields studied in the main text, $k_\Lambda = 0.025K_D$ suffices for the bilayer, while the trilayer requires $k_\Lambda = 0.035K_D$.\ It is worth noting that the size of the Fermi surface near the vHs, and hence the adequate $k_\Lambda$, increase with displacement field.

\section{Real space screened Coulomb potential}
\begin{figure}[h]
    \centering
    \includegraphics[width=.5\textwidth]{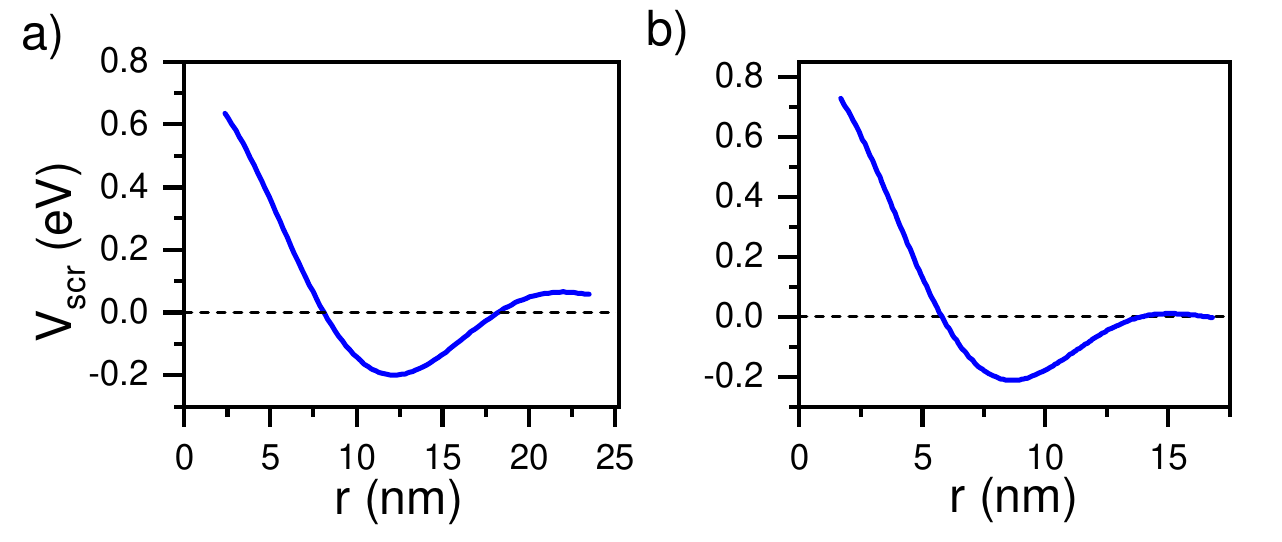}
    \caption{Real space cut along the $x$-direction of the screened Coulomb potential.\ (a) in BBG near the hole-doped vHs at $T=10$ mK.\ (b) in RTG near the hole-doped vHs at $T=65$ mK.}
    \label{fig:sp_RealSpace_ScreenedPot}
\end{figure}

A cut along the $x$-direction of the screened Coulomb potential in real space is shown in Fig.\ \ref{fig:sp_RealSpace_ScreenedPot}.\ For this, we compute the Fourier transform of the potential via, 
\begin{equation}
    V\left(\vb{r}\right) =\Delta_{\vb{k}}^{2}\sum_{\vb{k}} V_{scr}\left(\vb{k}\right) e^{-\mathrm{i}\vb{k}\cdot\vb{r}},
\end{equation}
where $\Delta_{\vb{k}}$ is the spacing between adjacent points in $\vb{k}$-space. 
As can be seen in Fig.\ \ref{fig:sp_RealSpace_ScreenedPot}, the repulsive Coulomb potential develops into an effective attractive potential once the screening is taken into consideration at a temperature around the $T_{c}$ of BBG and RTG.\ Since the continuum model yields the correct electronic properties up to a cutoff in the reciprocal space defined by $k_\Lambda$, we think that the position of the minimum of the screened potential is overestimated as the short-range information cannot be well captured with this model, and the magnitude is also unreliable.\ However, it is good enough to qualitatively describe the appearance of an attractive potential region which promotes superconductivity.

\section{Ising spin-orbit coupling}
\begin{figure}[h]
    \centering
    \includegraphics[width=.52\textwidth]{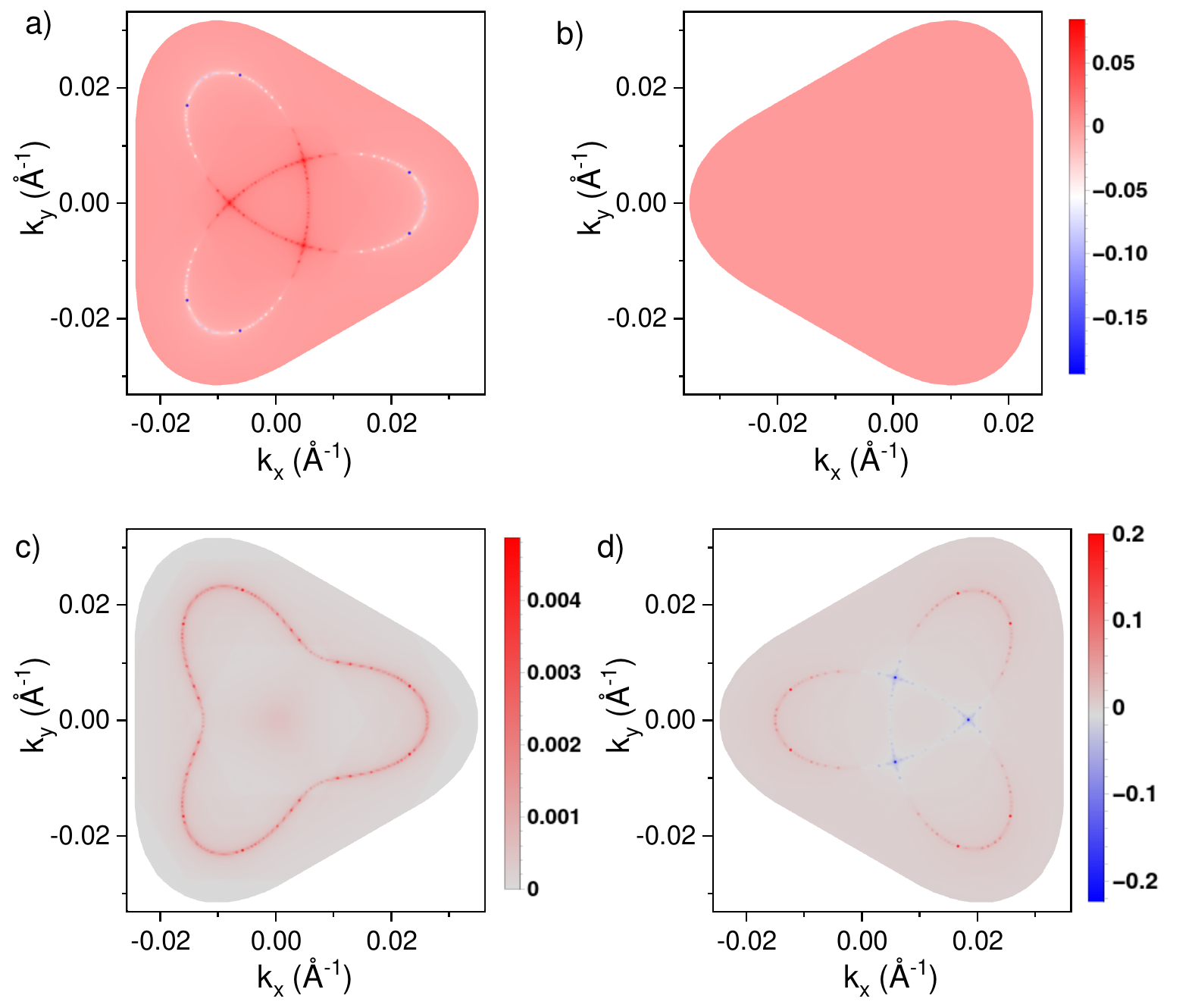}
    \caption{Order parameter of hole doped BBG with Ising SOC $\lambda_{I} = 1$ meV, at $\mu=\mu_{0}+\lambda_{I}$, in (a) in valley K+ and (b) in valley K-, (c,d) same as (a,b), at $\mu=\mu_{0}-\lambda_{I}$, where $\mu_{0}$ is the Fermi energy at which the vHs appears in absence of Ising SOC.}
    \label{fig:sp_orderPar_Ising}
\end{figure}

To account for the effects induced in the BBG by the proximity of WSe$_{2}$ and following the ideas introduced in \cite{zhang2022spin}, we include Ising spin-orbit coupling (SOC) to check if it can give rise to an enhancement of the critical temperature as measured experimentally.\ The Ising SOC results in an energy imbalance between Cooper pairs $\ket{\vb{K}^{+},\uparrow; \vb{K}^{-},\downarrow}$ and $\ket{\vb{K}^{+},\downarrow; \vb{K}^{-},\uparrow}$, making one of them to rise in energy by a quantity given by the SOC strength parameter, $\lambda_{I}$, while lowering the other by the same amount.\ Thus, the spin-valley-dependent hamiltonian for Ising SOC is given by, 

\begin{equation}
    \mathcal{H}_{I,\xi,s} =s\xi\lambda_{I}\mathbb{I},
    \label{eq:S3}
\end{equation}

where $\mathbb{I}$ is the identity matrix.\ As can be inferred from Eq.\ (\ref{eq:S3}) and seen in Figure \ref{fig:sp_orderPar_Ising}, Ising SOC locks the spin and valley degrees of freedom in the OP, making inapplicable the classification as spin-singlet or spin-triplet. 

\begin{figure}[h]
    \centering
    \includegraphics[width=\textwidth]{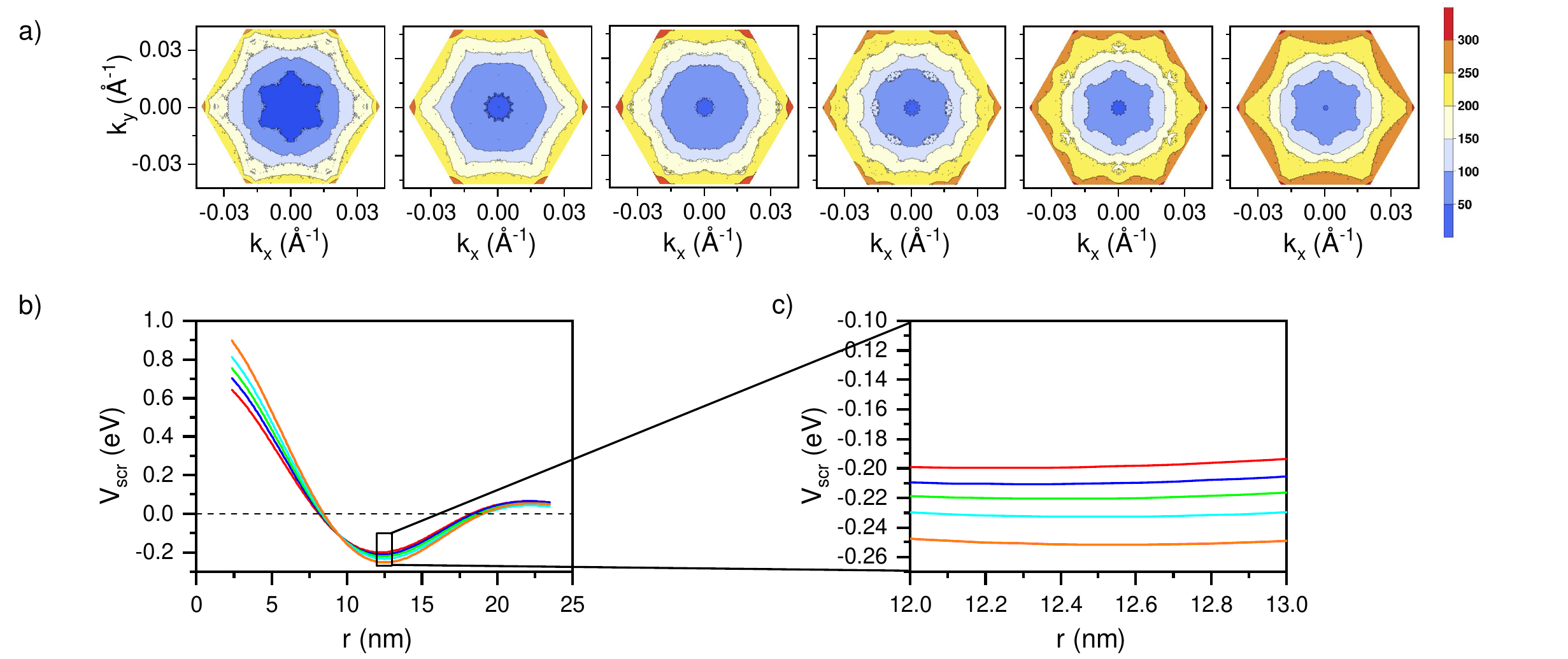}
    \caption{(a) Screened potential in BBG in reciprocal space near the hole-doped vHs at $\mu=\mu_0+\lambda_I$, at $T_{c}$ given in Fig.\ \textcolor{red}{3} in the main text, for $\lambda_{I} = 0,1,2,3,4,5$ meV, from left to right.\ (b) Real space screened potential for $\lambda_{I} = 0$ meV (red), $\lambda_{I} = 1$ meV (blue), $\lambda_{I} = 2$ meV (green), $\lambda_{I} = 3$ meV (cyan), $\lambda_{I} = 4$ meV (magenta), $\lambda_{I} = 5$ meV (orange).\ (c) Inset of (b) near the minimum of the screened potential.}
    \label{fig:sp_screenedPot_Ising}
\end{figure}

In the presence of SOC, the original vHs at $\mu_0$, which is four-fold degenerate, splits into two vHs which are two-fold degenerate, separated by $2\lambda_I$.\ A practical consequence is that the susceptibility in Eq.\ \textcolor{red}{3} in the main text needs to be separated in two parts, each with a prefactor of $N_f=2$, which are then summed to give the total susceptibility.\ Regarding the screening of the Coulomb potential, in reciprocal space, increasing the strength of Ising SOC narrows the region around the Dirac point where the minimum of the screened potential is concentrated, and makes it grow faster as a function of momentum.\ In real space this directly translates into a more attractive potential, see Fig.\ \ref{fig:sp_screenedPot_Ising}, which is consistent with the increment in critical temperature shown in Fig.\ \textcolor{red}{3} in the main text.\ The saturation of the critical temperature corresponds to the saturation of the minimum of the screened potential in real space, as reflected in Table \textcolor{red}{I}.

\begin{table}[h]
\begin{tabular}{|l|l|l|}
\hline
$\lambda_{I}$ (meV) & $r_{min}$ (nm) & $V_{scr}(r_{min})$ (eV) \\ \hline
0.0                 & 12.1985        & -0.199664               \\ \hline
1.0                 & 12.2834        & -0.210588               \\ \hline
2.0                 & 12.3683        & -0.220475               \\ \hline
3.0                 & 12.4532        & -0.232751               \\ \hline
4.0                 & 12.5381        & -0.251644               \\ \hline
5.0                 & 12.5381        & -0.251644               \\ \hline
\end{tabular}
\label{tab:table_isingSOC}
\caption{Relation between Ising SOC strength, $\lambda_{I}$, the minimum of the screened potential and its position in real space near the hole-doped vHs of BBG at the critical temperatures shown in Fig.\ \textcolor{red}{3} in the main text.}
\end{table}

\begin{figure}[h]
    \centering
    \includegraphics[width=.35\textwidth]{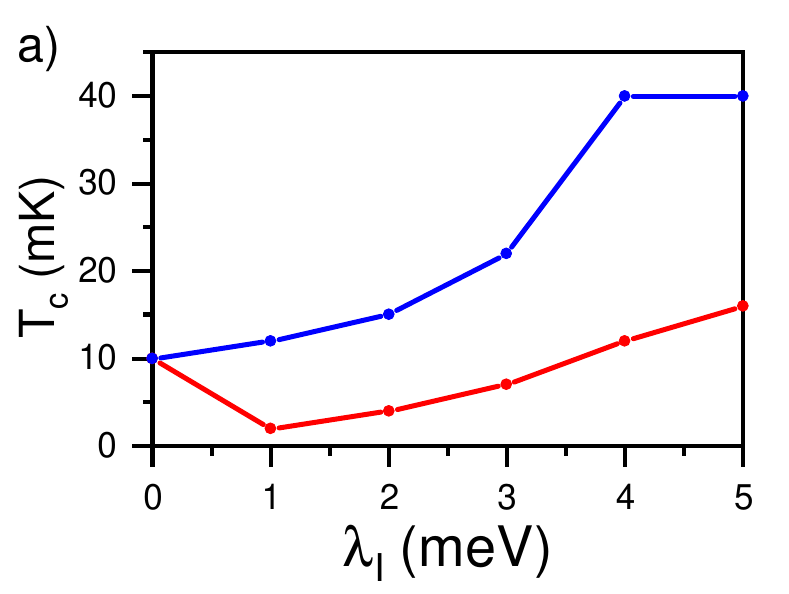}
    \caption{Critical temperature in hole-doped BBG with Ising spin-orbit coupling, at both vHs.\ Blue data correspond to the vHs at $\mu=\mu_{0}+\lambda_{I}$, while red data correspond to the vHs at $\mu=\mu_{0}-\lambda_{I}$, where $\mu_{0}$ is the Fermi energy at which the vHs appears in absence of Ising SOC.}
    \label{fig:sp_ising_criticalTemp_leftVHS}
\end{figure}

As for the results, the splitting induces two superconducting peaks as a function of Fermi energy, as shown in Fig.\ \ref{fig:sp_ising_criticalTemp_leftVHS} and seen in the experiments of Ref.\ \cite{zhang2022spin}.\ When the Fermi energy is placed at the $\mu_0+\lambda_I$ vHs, the vHs at $\mu_0-\lambda_I$ is nearly or completely unpopulated, depending on the value of $\lambda_I$, so the non-superconducting state is nearly or completely polarized to a half-metal.\ In this case, superconductivity is notably enhanced.\ In contrast, when $\mu$ is at the $\mu_0-\lambda_I$ vHs, the flavours at the $\mu_0+\lambda_I$ vHs have large Fermi surfaces.\ This situation leads to smaller critical temperatures than without spin-orbit, for any reasonable value of $\lambda_I$, which implies that the large Fermi surfaces away from $\mu$ have a detrimental effect on superconductivity.

Finally, it is also possible to consider a superconducting wavefunction involving only Cooper pairs $\ket{K^+\uparrow,K^-\downarrow}$, which interact with themselves via spin-valley flips.\ In this case, we find the same critical temperatures as shown in Fig.\ \textcolor{red}{3} in the main text for the vHs at $\mu_0+\lambda_I$.

\section{Superconductivity in BBG with electron doping}
\begin{figure}[h]
    \centering
    \includegraphics[width=\textwidth]{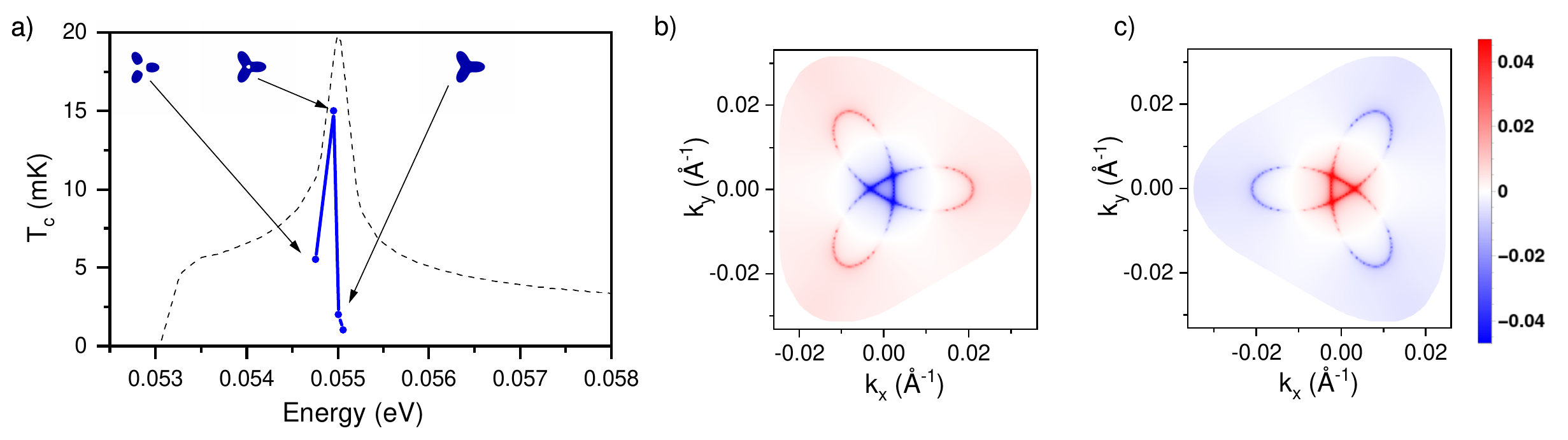}
    \caption{(a) Critical temperature (blue) and DOS (a.u., dashed black) versus Fermi energy in electron-doped BBG with a field-induced gap of 98 meV.\ The insets show the shape of the Fermi surface.\ (b) Order parameter at the energy for which $T_c=15$ mK, in valley $\vb{K}^{+}$ and (c) valley $\vb{K}^{-}$.\ Again a sign change between valleys is visible, thus pointing to the valley-singlet, spin-triplet nature of the superconducting state in BBG. The maximum $T_{c}$ is found near the vHs, with $n_e\approx0.385\cdot 10^{12}$ cm$^{-2}$.}
    \label{fig:sp_criticalTemp_BBGelectronSide}
\end{figure}

As can be seen in Fig.\ \textcolor{red}{2} in the main text, the critical temperature peaks close to the vHs in the DOS, which suggests that superconductivity could also exist near the vHs of the conduction band.\ In order to confirm this we compute the critical temperature around this vHs with the same procedure.\ The results can be found in Fig.\ \ref{fig:sp_criticalTemp_BBGelectronSide}.\ We find a $T_{c}\approx15$ mK near the vHs of the conduction band in BBG.\ After including the short-range interaction, the critical temperature increases to $T_{c}\approx25$ mK.\ The OP is qualitatively similar to the one with hole-doping.

\section{Diagrams}

\begin{figure}[h]
    \centering
    \includegraphics[width=.8\textwidth]{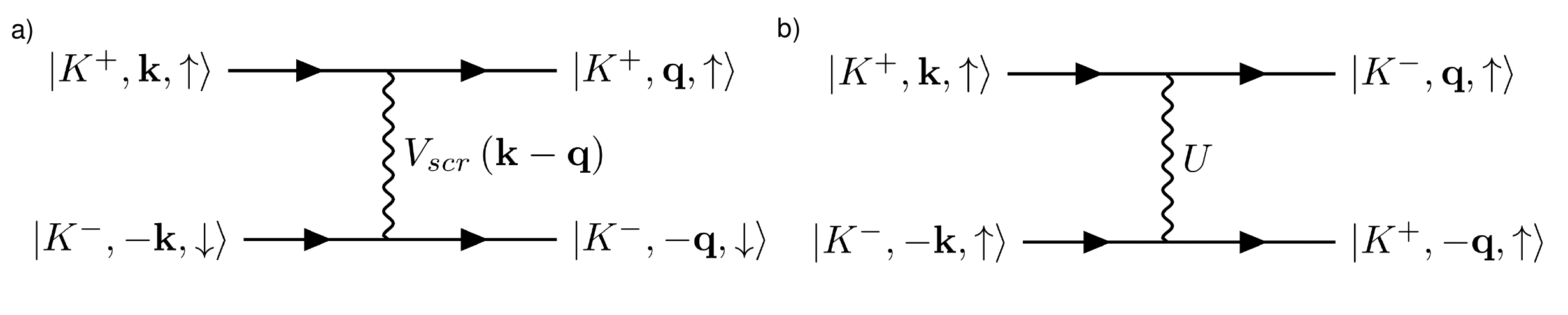}
    \caption{Diagrams of the (a) long-range interaction mediated by the screened Coulomb potential $V_{scr}$, which does not change the valley of the incoming states.\ (b) Short-range interaction mediated by the momentum independent Hubbard $U$, which makes the incoming states exchange valley.}
    \label{fig:diagrams}
\end{figure}

Figure \ref{fig:diagrams} shows the diagrams involved in superconductivity.\ We assume that Cooper pairs are formed between electrons in different valleys.\ Diagram (a) represents the long-range interaction, mediated by the screened Coulomb potential.\ After scattering, each carrier remains in its original valley.\ The kernel $\Gamma^+$ corresponds to this diagram, while $\Gamma^-$ corresponds to a diagram with $K^+$ and $K^-$ exchanged.\ The global wavefunction is $\ket{\Psi}\propto\ket{\vb{K}^{+},\uparrow; \vb{K}^{-},\downarrow}\pm \ket{\vb{K}^{+},\downarrow; \vb{K}^{-},\uparrow}$.\ The `+’ combination corresponds to valley-singlet, spin-triplet pairing, while the `–’ combination results in valley-triplet, spin-singlet pairing.\ In the absence of short-range interactions, these two pairings are degenerate.\ Diagram (b) represents the short-range interaction of the model, which is mediated by a repulsive, momentum-independent Hubbard $U=3$ eV.\ After scattering, the carriers exchange valley.\ This interaction can break the degeneracy between the two pairings mentioned above.\ 

Focusing on the superconducting OP in Fig.\ \textcolor{red}{4} in the main text, one might wonder why the OP in each panel lives in the Fermi surface of only one valley, even though one panel corresponds to $\tilde{\Delta}^+$ and the other to $\tilde{\Delta}^-$, each of which stems from diagrams involving both valleys.\ The reason for this is the application of time reversal symmetry (TRS).\ The two matrix elements involved in diagram (a) can be expressed in terms of quantities in valley $K^+$ only, after applying TRS.\ An analogous argument applies to the panel that plots the $\tilde{\Delta}^-$ part of the eigenvector. This time, we use TRS to express all quantities in the valley $K^-$.\ The short-range interaction is a small perturbation that does not change this picture.
\section{The gap equation}
We construct the gap equation following similar steps as if we were working with the BCS theory but instead of setting an electron-phonon coupling as mediator of the interaction between the pairs, we assume that the electron-electron interaction alone leads to pairing, which essentially is the Kohn-Luttinger methodology in \cite{Kohn1965}.\ This purely electronic interaction is the screened long-range Coulomb potential that a carrier feels within the system, which ultimately we consider to be the glue of the pairs.\ As there are many carriers playing a role in electronic interactions we account for the screening of the potential via the Random Phase Approximation (RPA).

Now we proceed to derive Eqs.\ (\textcolor{red}{6}, \textcolor{red}{7}) in the main text.\ We start from the gap equation (\textcolor{red}{4}), which includes two terms, one for the screened long-range Coulomb potential $V_{scr}$ and another for the short-range Hubbard repulsion $U$,
\begin{equation}
    \tilde{\Delta}^{+,i,j}\left(\vb{k}\right) = -\frac{K_{B}T}{\Omega}\sum_{\vb{q}, \omega}\sum_{i^{\prime},j^{\prime}}\left[  V_{scr}\left(\vb{k} - \vb{q}\right) G_{K^{+}}^{i,i^{\prime}}\left(\vb{q},\mathrm{i}\omega\right)G_{K^{-}}^{j,j^{\prime}}\left(-\vb{q},-\mathrm{i}\omega\right)\tilde{\Delta}^{+, i^{\prime},j^{\prime}}\left(\vb{q}\right) + U G_{K^{-}}^{i,i^{\prime}}\left(\vb{q},\mathrm{i}\omega\right)G_{K^{+}}^{j,j^{\prime}}\left(-\vb{q},-\mathrm{i}\omega\right)\tilde{\Delta}^{-, j^{\prime},i^{\prime}}\left(-\vb{q}\right)  \right].
    \label{eq:S4}
\end{equation}
We can work separately with the long-range and short-range terms and then simply develop the equations.\ We begin with the kernel of the long-range interaction and substitute the electrons' Green's functions by their definitions, $ G_{\vb{K}^{\xi}}^{i,i^{\prime}}\left(\vb{k},\pm\mathrm{i}\omega\right) = \sum_{n}\frac{\Psi^{i}_{\xi,n}\left(\vb{k}\right) \Psi^{i^{\prime},*}_{\xi,n}\left(\vb{k}\right) }{\mathrm{i}\omega \mp \epsilon_{\vb{K}^{\xi},n}\left(\vb{k}\right)}$, leading to,\\
\begin{equation}
\hspace*{-2.2cm}
    \begin{split}
        \tilde{\Delta}^{+,i,j}_{n,m}\left(\vb{k}\right) &= -\frac{K_{B}T}{\Omega}\sum_{\vb{q}, \omega}\sum_{i^{\prime},j^{\prime}} V_{scr}\left(\vb{k} - \vb{q}\right) G_{K^{+}}^{i,i^{\prime}}\left(\vb{q},\mathrm{i}\omega\right)G_{K^{-}}^{j,j^{\prime}}\left(-\vb{q},-\mathrm{i}\omega\right)\tilde{\Delta}^{+, i^{\prime},j^{\prime}}\left(\vb{q}\right),\\
        &=-\frac{K_{B}T}{\Omega}\sum_{\vb{q}, \omega}\sum_{i^{\prime},j^{\prime}} V_{scr}\left(\vb{k} - \vb{q}\right)\frac{\Psi^{i}_{K^{+},n}\left(\vb{q}\right) \Psi^{i^{\prime},*}_{K^{+},n}\left(\vb{q}\right) }{\mathrm{i}\omega - \epsilon_{K^{+},n}\left(\vb{q}\right)} \frac{\Psi^{j}_{K^{-},m}\left(-\vb{q}\right) \Psi^{j^{\prime},*}_{K^{-},m}\left(-\vb{q}\right) }{\mathrm{i}\omega + \epsilon_{K^{-},m}\left(-\vb{q}\right)}\tilde{\Delta}^{+, i^{\prime},j^{\prime}}\left(\vb{q}\right),\\
        &=-\frac{K_{B}T}{\Omega}\sum_{\vb{q}} V_{scr}\left(\vb{k} - \vb{q}\right)\Psi^{i}_{K^{+},n}\left(\vb{q}\right)\Psi^{j}_{K^{-},m}\left(-\vb{q}\right)\sum_{\omega}\frac{1}{\mathrm{i}\omega - \epsilon_{K^{+},n}\left(\vb{q}\right)}\frac{1}{\mathrm{i}\omega + \epsilon_{K^{-},m}\left(-\vb{q}\right)}\sum_{i^{\prime},j^{\prime}}  \Psi^{i^{\prime},*}_{K^{+},n}\left(\vb{q}\right)  \Psi^{j^{\prime},*}_{K^{-},m}\left(-\vb{q}\right)\tilde{\Delta}^{+,i^{\prime},j^{\prime}}\left(\vb{q}\right),\\
        &=-\frac{K_{B}T}{\Omega}\sum_{\vb{q}} V_{scr}\left(\vb{k} - \vb{q}\right)\Psi^{i}_{n}\left(\vb{q}\right)\Psi^{j,*}_{m}\left(\vb{q}\right)\sum_{\omega} \frac{1}{\mathrm{i}\omega - \epsilon_{n}\left(\vb{q}\right)}\frac{1}{\mathrm{i}\omega + \epsilon_{m}\left(\vb{q}\right)}\sum_{i^{\prime},j^{\prime}}  \Psi^{i^{\prime},*}_{n}\left(\vb{q}\right)  \Psi^{j^{\prime}}_{m}\left(\vb{q}\right)\tilde{\Delta}^{+, i^{\prime},j^{\prime}}\left(\vb{q}\right),\\
        &=-\frac{1}{\Omega}\sum_{\vb{q}} V_{scr}\left(\vb{k} - \vb{q}\right)\Psi^{i}_{n}\left(\vb{q}\right)\Psi^{j,*}_{m}\left(\vb{q}\right)\frac{f(-\epsilon_{m}\left(\vb{q}\right)) - f(\epsilon_{n}\left(\vb{q}\right))}{\epsilon_{n}\left(\vb{q}\right)+\epsilon_{m}\left(\vb{q}\right)}\tilde{\Delta}_{n,m}^{+}\left(\vb{q}\right),
    \end{split}
    \label{eq:S6}
\end{equation}
where in the last step we have performed a Matsubara sum using $\sum_{\mathrm{i}\omega}\frac{1}{\mathrm{i}\omega - \epsilon_{1}}\frac{1}{\mathrm{i}\omega + \epsilon_{2}} = \frac{1}{K_{B} T}\frac{1}{\epsilon_{1}+\epsilon_{2}}\left[f(-\epsilon_{2}) - f(\epsilon_{1}) \right]$, thus,
\begin{equation}
    \sum_{i,j}\Psi^{i,*}_{n}\left(\vb{k}\right)  \Psi^{j}_{m}\left(\vb{k}\right)\tilde{\Delta}^{+,i,j}_{n,m}\left(\vb{k}\right) = -\frac{1}{\Omega}\sum_{\vb{q}} V_{scr}\left(\vb{k} - \vb{q}\right)\frac{f(-\epsilon_{m}\left(\vb{q}\right)) - f(\epsilon_{n}\left(\vb{q}\right))}{\epsilon_{n}\left(\vb{q}\right)+\epsilon_{m}\left(\vb{q}\right)}\sum_{i,j}\Psi^{i,*}_{n}\left(\vb{k}\right)  \Psi^{j}_{m}\left(\vb{k}\right)\Psi^{i}_{n}\left(\vb{q}\right)\Psi^{j,*}_{m}\left(\vb{q}\right)\tilde{\Delta}_{n,m}^{+}\left(\vb{q}\right).
    \label{eq:S7}
\end{equation}
Here we introduce the new quantity $\tilde{\tilde{\Delta}}^{+}\left(\vb{p}\right)=\tilde{\Delta}^{+}\left(\vb{p}\right) \sqrt{\frac{f(-\epsilon_{m}\left(\vb{p}\right)) - f(\epsilon_{n}\left(\vb{p}\right))}{\epsilon_{n}\left(\vb{p}\right)+\epsilon_{m}\left(\vb{p}\right)}}$ for the kernel to be hermitian, so after the substitution,
\begin{equation}
\hspace*{-1cm}
    \tilde{\tilde{\Delta}}_{n,m}^{+}\left(\vb{k}\right) = -\frac{1}{\Omega}\sum_{\vb{q}} V_{scr}\left(\vb{k} - \vb{q}\right) \sqrt{\frac{f(-\epsilon_{m}\left(\vb{k}\right)) - f(\epsilon_{n}\left(\vb{k}\right))}{\epsilon_{n}\left(\vb{k}\right)+\epsilon_{m}\left(\vb{k}\right)}} \sqrt{\frac{f(-\epsilon_{m}\left(\vb{q}\right)) - f(\epsilon_{n}\left(\vb{q}\right))}{\epsilon_{n}\left(\vb{q}\right)+\epsilon_{m}\left(\vb{q}\right)}}\sum_{i,j}\Psi^{i,*}_{n}\left(\vb{k}\right)  \Psi^{j}_{m}\left(\vb{k}\right)\Psi^{i}_{n}\left(\vb{q}\right)\Psi^{j,*}_{m}\left(\vb{q}\right)\tilde{\Delta}^{+}_{n,m}\left(\vb{q}\right).
    \label{eq:S8}
\end{equation}
Now, following the same procedure we continue with the kernel of the short-range interaction.
\begin{equation}
    \begin{split}
        \tilde{\Delta}^{+,i,j}_{n,m}\left(\vb{k}\right) &= -U\frac{K_{B}T}{\Omega}\sum_{\vb{q}, \omega}\sum_{i^{\prime},j^{\prime}} G_{K^{-}}^{i,i^{\prime}}\left(\vb{q},\mathrm{i}\omega\right)G_{K^{+}}^{j,j^{\prime}}\left(-\vb{q},-\mathrm{i}\omega\right)\tilde{\Delta}^{-, j^{\prime},i^{\prime}}\left(-\vb{q}\right) ,\\
        &=-U\frac{K_{B}T}{\Omega}\sum_{\vb{q}, \omega}\sum_{i^{\prime},j^{\prime}} \frac{\Psi^{i}_{K^{-},n}\left(\vb{q}\right) \Psi^{i^{\prime},*}_{K^{-},n}\left(\vb{q}\right) }{\mathrm{i}\omega - \epsilon_{K^{-},n}\left(\vb{q}\right)} \frac{\Psi^{j}_{K^{+},m}\left(-\vb{q}\right) \Psi^{j^{\prime},*}_{K^{+},m}\left(-\vb{q}\right) }{\mathrm{i}\omega + \epsilon_{K^{+},m}\left(-\vb{q}\right)}\tilde{\Delta}^{-, j^{\prime},i^{\prime}}\left(-\vb{q}\right),\\
        &=-U\frac{K_{B}T}{\Omega}\sum_{\vb{q}}
        \Psi^{i}_{K^{-},n}\left(\vb{q}\right)\Psi^{j}_{K^{+},m}\left(-\vb{q}\right) \sum_{\omega}\frac{1}{\mathrm{i}\omega - \epsilon_{K^{-},n}\left(\vb{q}\right)}\frac{1}{\mathrm{i}\omega + \epsilon_{K^{+},m}\left(-\vb{q}\right)}\sum_{i^{\prime},j^{\prime}}  \Psi^{i^{\prime},*}_{K^{-},n}\left(\vb{q}\right) \Psi^{j^{\prime},*}_{K^{+},m}\left(-\vb{q}\right)\tilde{\Delta}^{-, j^{\prime},i^{\prime}}\left(-\vb{q}\right),\\
        &=-U\frac{K_{B}T}{\Omega}\sum_{\vb{q}} \Psi^{i,*}_{n}\left(-\vb{q}\right)\Psi^{j}_{m}\left(-\vb{q}\right)\sum_{\omega}\frac{1}{\mathrm{i}\omega - \epsilon_{n}\left(-\vb{q}\right)}\frac{1}{\mathrm{i}\omega + \epsilon_{m}\left(-\vb{q}\right)}\sum_{i^{\prime},j^{\prime}} \Psi^{i^{\prime}}_{n}\left(-\vb{q}\right) \Psi^{j^{\prime},*}_{m}\left(-\vb{q}\right)\tilde{\Delta}^{-, j^{\prime},i^{\prime}}\left(-\vb{q}\right),\\
        &=-U\frac{K_{B}T}{\Omega}\sum_{\vb{q}} \Psi^{i,*}_{n}\left(\vb{q}\right)\Psi^{j}_{m}\left(\vb{q}\right)\sum_{\omega}\frac{1}{\mathrm{i}\omega - \epsilon_{n}\left(\vb{q}\right)}\frac{1}{\mathrm{i}\omega + \epsilon_{m}\left(\vb{q}\right)}\sum_{i^{\prime},j^{\prime}} \Psi^{i^{\prime}}_{n}\left(\vb{q}\right) \Psi^{j^{\prime},*}_{m}\left(\vb{q}\right)\tilde{\Delta}^{-, j^{\prime},i^{\prime}}\left(\vb{q}\right),\\
        &=-U\frac{K_{B}T}{\Omega}\sum_{\vb{q}} \Psi^{i,*}_{n}\left(\vb{q}\right)\Psi^{j}_{m}\left(\vb{q}\right)\sum_{\omega}\frac{1}{\mathrm{i}\omega - \epsilon_{n}\left(\vb{q}\right)}\frac{1}{\mathrm{i}\omega + \epsilon_{m}\left(\vb{q}\right)}\sum_{i^{\prime},j^{\prime}}\Psi^{i^{\prime},*}_{m}\left(\vb{q}\right)\Psi^{j^{\prime}}_{n}\left(\vb{q}\right)\tilde{\Delta}^{-, i^{\prime},j^{\prime}}\left(\vb{q}\right),\\
        &=-U\frac{1}{\Omega}\sum_{\vb{q}} \Psi^{i,*}_{n}\left(\vb{q}\right)\Psi^{j}_{m}\left(\vb{q}\right)\frac{f(-\epsilon_{m}\left(\vb{q}\right)) - f(\epsilon_{n}\left(\vb{q}\right))}{\epsilon_{n}\left(\vb{q}\right)+\epsilon_{m}\left(\vb{q}\right)}\tilde{\Delta}_{n,m}^{-}\left(\vb{q}\right),
    \end{split}
    \label{eq:S9}
\end{equation}
thus,
\begin{equation}
    \sum_{i,j}\Psi^{i,*}_{n}\left(\vb{k}\right) \Psi^{j}_{m}\left(\vb{k}\right)\tilde{\Delta}^{+,i,j}_{n,m}\left(\vb{k}\right) = -U\frac{1}{\Omega}\sum_{\vb{q}}\frac{f(-\epsilon_{m}\left(\vb{q}\right)) - f(\epsilon_{n}\left(\vb{q}\right))}{\epsilon_{n}\left(\vb{q}\right)+\epsilon_{m}\left(\vb{q}\right)}\sum_{i,j}\Psi^{i,*}_{n}\left(\vb{k}\right)  \Psi^{j}_{m}\left(\vb{k}\right)\Psi^{i,*}_{n}\left(\vb{q}\right)\Psi^{j}_{m}\left(\vb{q}\right)\tilde{\Delta}_{n,m}^{-}\left(\vb{q}\right),
    \label{eq:S10}
\end{equation}
and we arrive at,
\begin{equation}
    \tilde{\Delta}_{n,m}^{+}\left(\vb{k}\right) = -U\frac{1}{\Omega}\sum_{\vb{q}} \sqrt{\frac{f(-\epsilon_{m}\left(\vb{k}\right)) - f(\epsilon_{n}\left(\vb{k}\right))}{\epsilon_{n}\left(\vb{k}\right)+\epsilon_{m}\left(\vb{k}\right)}} \sqrt{\frac{f(-\epsilon_{m}\left(\vb{q}\right)) - f(\epsilon_{n}\left(\vb{q}\right))}{\epsilon_{n}\left(\vb{q}\right)+\epsilon_{m}\left(\vb{q}\right)}}\sum_{i,j}\Psi^{i,*}_{n}\left(\vb{k}\right)  \Psi^{j}_{m}\left(\vb{k}\right)\Psi^{i,*}_{n}\left(\vb{q}\right)\Psi^{j}_{m}\left(\vb{q}\right)\tilde{\Delta}_{n,m}^{-}\left(\vb{q}\right).
    \label{eq:S11}
\end{equation}

Combining Eqs.\ (\ref{eq:S8}, \ref{eq:S11}) along with time reversal symmetry gives Eqs.\ (\textcolor{red}{6}, \textcolor{red}{7}) in the main text.

\section{Critical temperature dependence on electric field.}
\begin{figure}[h]
    \centering
    \includegraphics[width=.5\textwidth]{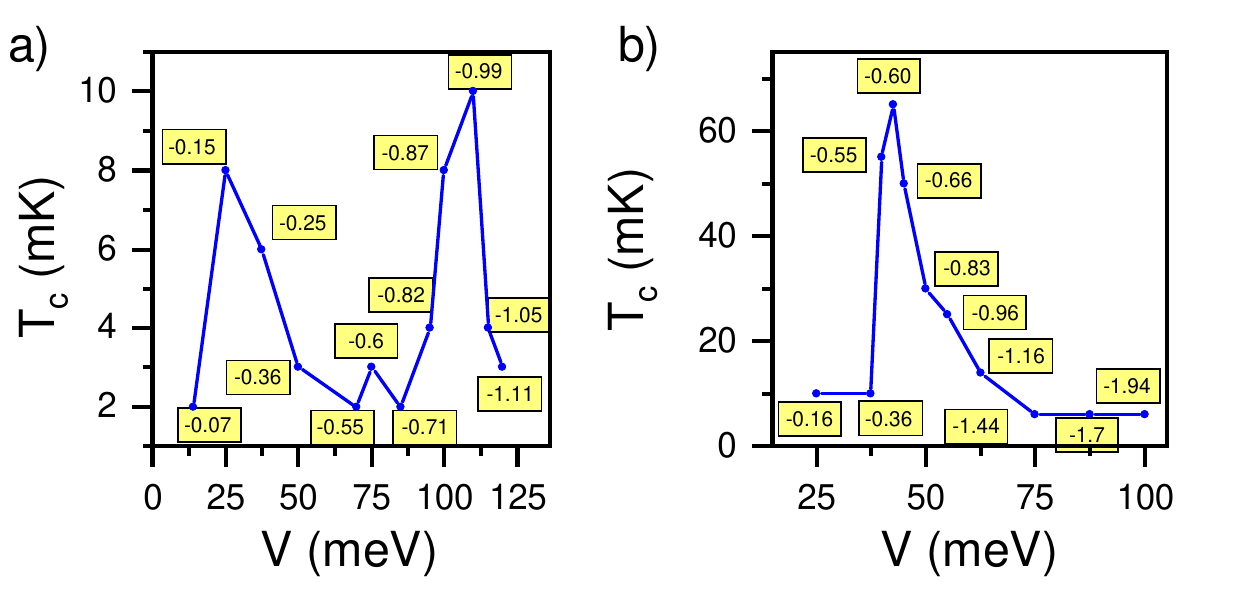}
    \caption{Electric field dependence of the critical temperature near the vHs of hole-doped (a) BBG and (b) RTG.\ Yellow boxes label the electronic density at which the critical temperature has been computed for each case, in units of $10^{12}$ cm$^{-2}$}
    \label{fig:sp_DispFieldDepen}
\end{figure}

Figure \ref{fig:sp_DispFieldDepen} indicates the critical temperatures of BBG and RTG, as a function of the external electric field.\ $T_c$ peaks at a certain field, and tends to diminish when the field changes.\ These results are in qualitative agreement with the experimental findings in Refs.\ \cite{zhou2021superconductivity,zhou2022isospin}.

\section{The role of the Hubbard U.}

\begin{figure}[h]
    \centering
    \includegraphics[width=\textwidth]{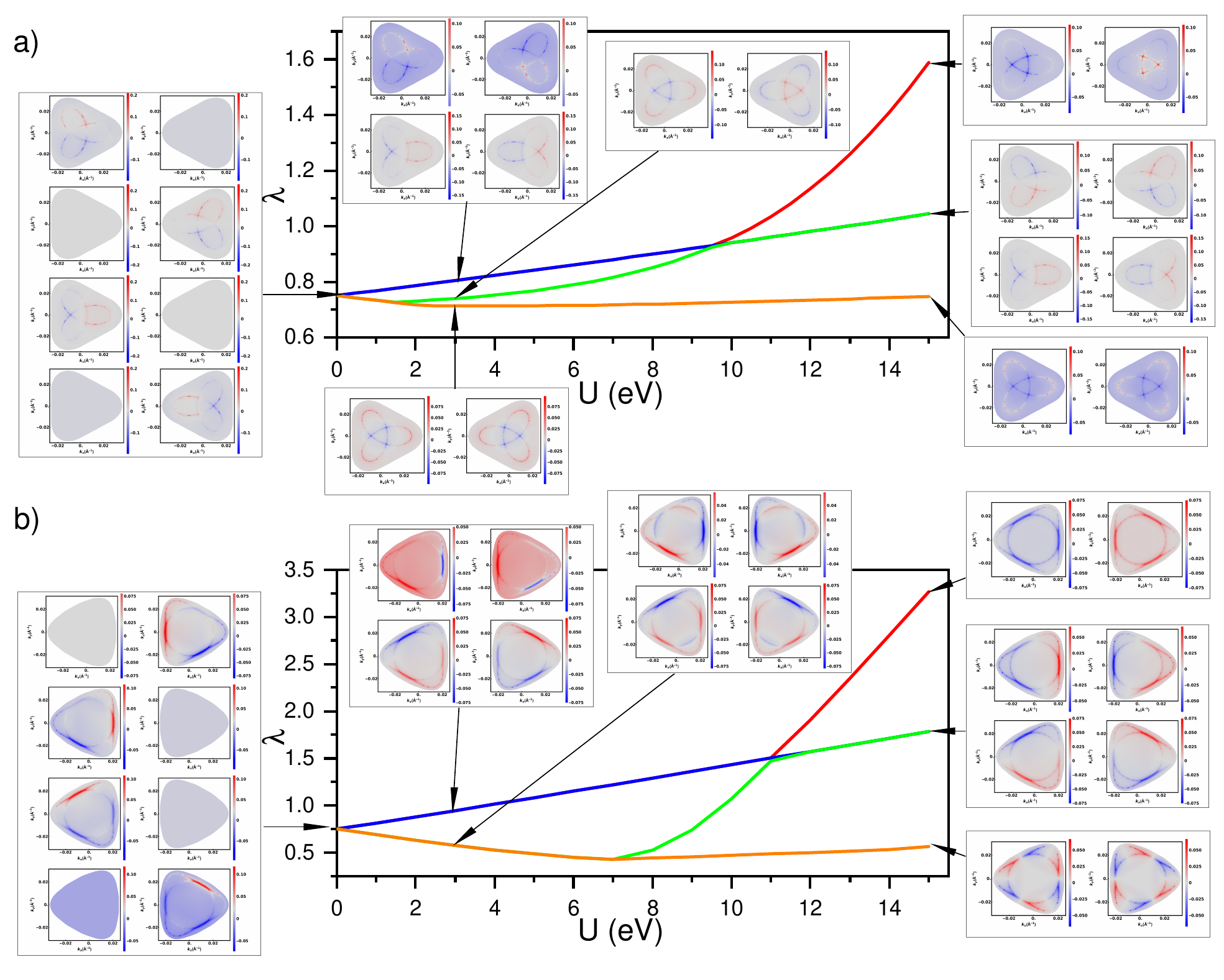}
    \caption{Symmetry of the OP, as a function of the short-range interaction strength, $U$, for the four largest eigenvalues near the hole-doped vHs of (a) BBG at $T=10$ mK and (b) RTG at $T=85$mK.\ These plots are meant to illustrate the form of the order parameter as a function of U, but note that the eigenvalues $\lambda<1$ imply that the superconducting state has not been reached here, and details change once it is, e.g.\ the degeneracy and associated $C_3$-symmetry breaking of the largest eigenvalue of BBG at $U=3$ eV disappears.}
    \label{fig:sp_BrokenDeg_Uvalue}
\end{figure}

We conclude with a tentative discussion on the effects of the value of $U$, which is the least well determined parameter in our model.\ Realistic values $U\sim 2-4$ eV.\ \cite{Wehling2011,Wehling2013}.\ We recall here that the screened long-range Coulomb interaction leads to a degeneracy between valley-triplet, spin-singlet and valley-singlet, spin-triplet pairings.\ The short-range Hubbard-like interaction can break this degeneracy, selecting a preferred pairing symmetry.\ For the value used in this work, $U=3$ eV, the short-range interaction can be seen as a small perturbation.\ In BBG and RTG, an infinitesimally small value of $U$ is enough to break the aforementioned degeneracy.\ To first order in perturbation theory, the change in the kernel eigenvalue due to the short-range interaction is:
\begin{equation}
\delta \epsilon=-\sum_{\vb{k},\vb{k}^{\prime}}\tilde{\Delta}_{\vb{k}} \tilde{U}(\vb{k},\vb{k}^{\prime}) \tilde{\Delta}_{\vb{k}^{\prime}},
\label{eq:perturbation}
\end{equation}
where $\tilde{\Delta}_{\vb{k}}$ and $\tilde{\Delta}_{\vb{k}^{\prime}}$ are the order parameters in valley $K^+$ and $K^-$, in the absence of the short-range interaction.\ From this follows the common wisdom that a repulsive, i.e.\ positive, U, favours valley-singlet pairing, in which the OP changes sign between valleys, so that $\delta \epsilon >0$, resulting in an enhanced critical temperature.\ However, Eq.\ \ref{eq:perturbation} is valid only for single pocket superconductors, while in BBG and RTG each valley shows at least two pockets, which makes the situation more complex, as pockets with a certain sign in one valley can interact strongly with others with the opposite sign in the other valley, even if the global state is valley-triplet, spin-singlet.\ Still, both BBG and RTG have valley-singlet, spin-triplet pairing for any finite value of U, which is the usual result with repulsive short-range interactions, as mentioned above.

It is illustrative to note that in the limit of large U, the short-range interaction is not a perturbation anymore, but rather becomes the leading interaction and overshadows the long-range Coulomb potential.\ When short-range interactions are dominant, the most energetically favorable solution is the one in which each valley as a constant sign, and there is a change of sign between valleys.\ Thus, for large values of $U$ the order parameter losses the sign changes within each valley.\ This phenomenon occurs in both materials, although in the bilayer the intravalley signs only disappear completely at $U\gtrsim 25$ eV.\ 

\section{Critical temperature starting from a Half Metal state.}

\begin{figure}[h]
    \centering
    \includegraphics[width=.9\textwidth]{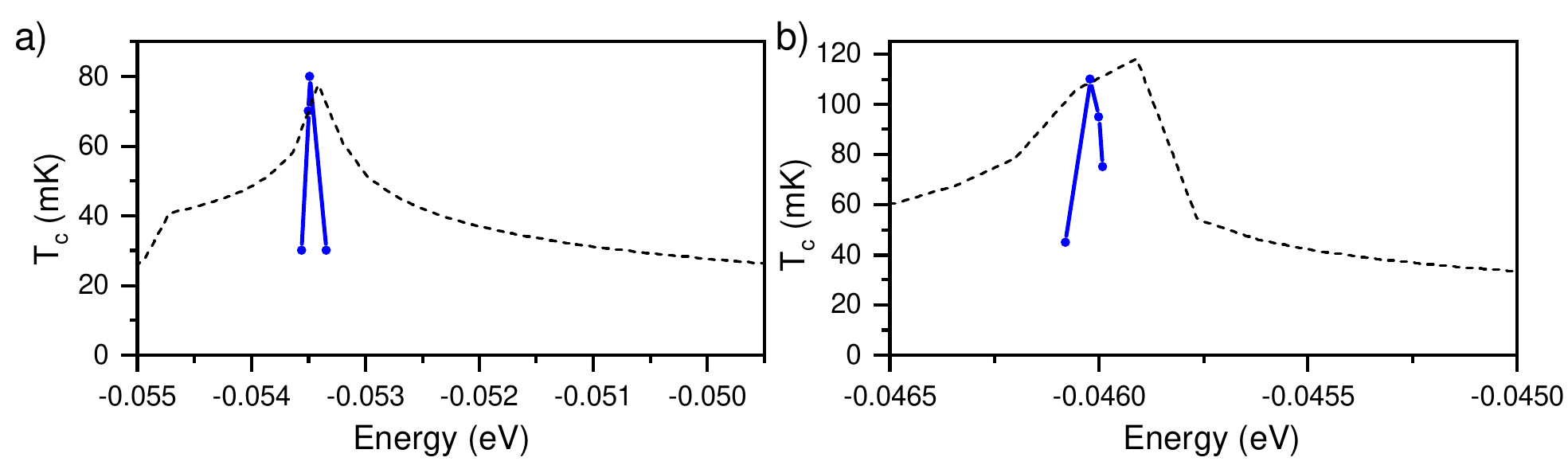}
    \caption{Superconducting critical temperature starting from a half metal non-superconducting state (solid blue) and DOS (a.u. dashed black) versus Fermi energy in (a) BBG with hole doping and a gap of $98$ meV; near the vHs we observe a $T_{c}\approx 80$ mK, and in (b) RTG with hole doping and a gap of $74$ meV; near the vHs we observe a $T_{c}\approx 110$ mK.}
    \label{fig:sp_Halfmetal}
\end{figure}

Figure \ref{fig:sp_Halfmetal} shows the critical temperature as a function of Fermi energy in both materials, when the non-superconducting state is a half metal.\ The calculation is equivalent to the one in Fig.\ \textcolor{red}{2} in the main text, but the factor of $N_f=4$ in the susceptibility is substituted by $N_f=2$.\ The situation is qualitatively similar to the one with Ising SOC.\ However, the complete disappearance of two flavours leads to a more severe change in the susceptibility than the energy split induced by Ising SOC. It is an approximation which gives a rough idea of what happens for very large values of $\lambda_I$.\ The critical temperature we find here for BBG is 80 mK, while starting from the full metal it is 10 mK, and in the experiment, 26 mK \cite{zhou2022isospin}.\ For RTG. the critical temperatures starting from a half metal or a full metal state are 110 mK and 65 mK, respectively, while the experiment reports 105 mK \cite{zhou2021superconductivity}.

\end{document}